\documentclass[twocolumn,english,pra,showpacs,superscriptaddress,longbibliography]{revtex4-1}
\usepackage{amsfonts}
\usepackage{amsmath}
\usepackage{amssymb}
\usepackage[colorlinks, citecolor=blue,linkcolor=red]{hyperref}
\usepackage{color}
\usepackage{float}
\usepackage{hyperref}
\usepackage{textcomp}
\usepackage[rightcaption]{sidecap}
\usepackage{subfigure}
\usepackage[utf8]{inputenc}
\usepackage[T1]{fontenc}
\usepackage[rightcaption]{sidecap}
\usepackage{graphicx}
\begin{document}
\title{$\mathcal{PT}$-symmetric cavity magnomechanics with gain-assisted transparency and amplification}
\author{Cham Oumie}
\affiliation{Department of Physics, Zhejiang Normal University, Jinhua 321004, China.}
\author{Wu-Ming Liu}
\email{wliu@iphy.ac.cn}
\affiliation{Beijing National Laboratory for Condensed Matter Physics, Institute of Physics, Chinese Academy of Sciences, Beijing 100190, China}
\author{Kashif Ammar Yasir}
\email{kayasir@zjnu.edu.cn}\affiliation{Department of Physics, Zhejiang Normal University, Jinhua 321004, China.}
\affiliation{Zhejiang Institute of Photoelectronics, Jinhua 321004, China.}
\setlength{\parskip}{0pt}
\setlength{\belowcaptionskip}{-10pt}
\begin{abstract}
We investigate magnomechanically induced transparency in a parity-time-symmetric cavity magnomechanical system with traveling-field–induced non-Hermiticity. The setup consists of a microwave cavity mode coupled to magnons in a single-crystal yttrium iron garnet sphere, which in turn are hybridized with a vibrational mechanical mode through magnetostrictive interaction. In the Hermitian regime, strong photon–magnon coupling generates a single transparency window in the cavity transmission, which splits into a doublet when the magnon is coherently hybridized with the mechanical mode via magnomechanical coupling. This establishes a versatile platform in which the transparency spectrum can be engineered from single- to multi-window response using experimentally accessible, scaled magnomechanical interactions. When a non-Hermitian coupling is introduced, the system enters a parity-time-broken regime in which the transparency ceases to be purely passive and becomes gain assisted, leading to asymmetric transmission with amplification on one side of the resonance and enhanced absorption on the other. By tuning the cavity detuning, we convert magnomechanical transparency into Fano-type line shapes with strongly non-Lorentzian phase dispersion and map their deformation into asymmetric, gain-assisted Fano ridges in the joint space of probe and magnon detunings. Finally, we analyze the associated group delay and show that both slow- and fast-light behavior can be widely tuned by varying the photon–magnon and magnomechanical couplings together with the non-Hermitian strength, highlighting parity-time-symmetric cavity magnomechanics as a promising platform for reconfigurable quantum signal processing and enhanced sensing.
\end{abstract}

\date{\today}
\maketitle

\section{Introduction}
Hybrid quantum systems that coherently couple magnons, photons, and phonons have emerged as versatile platforms for probing macroscopic quantum phenomena and realizing advanced quantum technologies~\cite{Aspelmeyer2014, Zhang2016_SciAdv, Zhang2016, Fan2022}. Among these, \emph{cavity magnomechanics}—in which a ferrimagnetic yttrium iron garnet (YIG) sphere interacts simultaneously with a microwave cavity field and its intrinsic mechanical vibrations—provides a distinctive route to quantum transduction, coherent state transfer, and nonclassical state generation~\cite{Hu2013, Tabuchi2015, Harder2021}. In such hybrid platforms, the magnetic dipole interaction mediates photon–magnon coupling, while magnetostrictive strain enables magnon–phonon coupling, giving rise to a tripartite system capable of achieving strong and even ultrastrong coupling regimes~\cite{Wang2018, Rao2020, Luo2023}.

The incorporation of \emph{non-Hermitian physics} into cavity magnomechanics has opened new avenues for controlling coherence and energy flow via engineered gain and loss~\cite{ElGanainy2018, Feng2017, Ozdemir2019}. In this framework, balanced amplification and dissipation establish \emph{parity–time (PT) symmetry}, allowing real eigenvalue spectra and unconventional dynamics even in open systems~\cite{Bender1998, Miri2019}. Non-Hermitian interactions naturally host \emph{exceptional points} (EPs)—spectral degeneracies at which eigenvalues and eigenvectors coalesce—leading to enhanced mode hybridization, asymmetric energy transfer, and amplified quantum responses~\cite{Peng2014, Zhang2019, Qian2024}. The exploration of EPs has unveiled striking effects including topological mode exchange, nonreciprocal phonon lasing, and ultrasensitive magnetometry~\cite{Jing2014, Zhang2020, Dai2024}.

Recently, non-Hermiticity has been \emph{actively engineered} in magnomechanical systems by introducing an additional traveling microwave or optical field that directly interacts with the YIG sphere~\cite{Chengyong2025}. This auxiliary field functions as a tunable gain channel, compensating intrinsic losses and enabling PT-symmetric operation. By adjusting its intensity and incidence angle, one can continuously tune the gain–loss balance, driving transitions between Hermitian, PT-symmetric, and broken-symmetry regimes. Such spatially directed gain control can even generate higher-order EPs and hybrid steady states in which magnons, photons, and phonons exhibit distinct coherence and stability features~\cite{Chengyong2025, Lai2024}. These advances demonstrate that spatially engineered energy injection offers a powerful means to manipulate topology, amplification, and coherence in hybrid quantum systems.

Parallel developments in \emph{quantum nonlinear optics} provide a conceptual foundation for understanding interference-based transparency phenomena~\cite{Fleischhauer2005}. A prime example is \emph{electromagnetically induced transparency} (EIT), in which destructive quantum interference between excitation pathways renders an opaque medium transparent to a probe field. Its mechanical counterpart, \emph{optomechanically induced transparency} (OMIT), has enabled slow light, steep dispersion, and quantum state storage in various physical systems~\cite{Weis2010, Safavi2011, Agarwal2010}. Extending this concept to hybrid magnonic systems yields \emph{magnomechanically induced transparency} (MMIT), a phenomenon arising from coherent photon–magnon–phonon coupling~\cite{Wang2018, Rao2020, Luo2023}. MMIT enables controllable transparency and absorption at microwave frequencies, forming a cornerstone for quantum communication and precision sensing.

Introducing non-Hermiticity into such hybrid platforms profoundly modifies their interference physics~\cite{Xu2022, He2019}. Gain–loss coupling and EP-induced mode coalescence reshape the transparency spectrum, giving rise to \emph{non-Hermiticity-induced transparency} with asymmetric transmission, amplified response, and tunable Fano profiles~\cite{Jing2014, Zhang2021, Yu2023}. The interplay between EIT-like coherence and PT-symmetric gain–loss control allows dynamical manipulation of quantum amplification, dissipation-induced coherence revival, and extreme phase dispersion near EPs~\cite{Yang2023, Ren2024, Qiu2023}. 

\emph{In this work}, we analyze a cavity magnomechanical architecture in which a ferrimagnetic magnon mode, hosted by a YIG sphere and coupled to both a microwave cavity field and a mechanical vibration, is rendered non-Hermitian by an additional traveling drive that supplies controllable magnonic gain. By adjusting the gain strength, the coherent coupling rates, and the relevant detunings, the system can be steered continuously from conventional magnomechanically induced transparency to regimes featuring amplified and strongly asymmetric transparency. The resulting modifications in phase dispersion and mechanical backaction provide a unified picture of transparency and amplification governed by engineered gain–loss, and outline realistic pathways toward tunable quantum amplifiers, slow- and fast-light control, and hybrid sensing devices operating in the vicinity of exceptional points.

The remainder of this paper is organized as follows. In Sec.~\ref{sec1} we introduce the hybrid cavity magnomechanical setup and present the corresponding theoretical model. In Sec.~\ref{sec2} we analyze the emergence of non-Hermitian MMIT and its dependence on the system parameters. Section~\ref{sec3} is devoted to the appearance and control of Fano resonances. In Sec.~\ref{sec4} we investigate the dynamics of slow and fast light in the presence of engineered gain and loss. Finally, Sec.~\ref{sec5} summarizes our main results and outlines possible directions for future work.
\section{System Description}\label{sec1}
\begin{figure}[tp]
	\includegraphics[width=7.5cm]{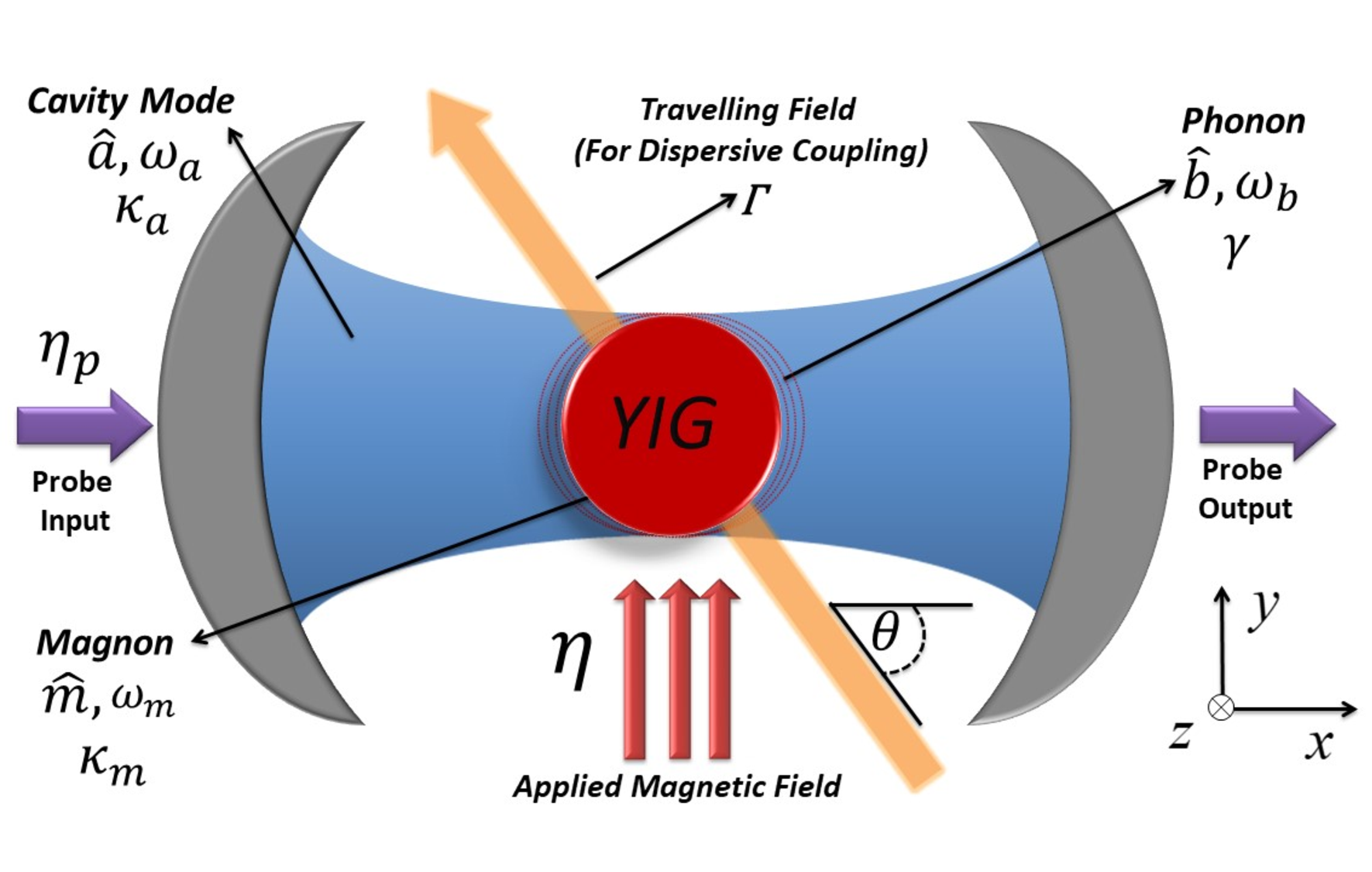}
	\caption{The schematic description of a diagram of the non-hermitian cavity magnomechanical system centered around a YIG sphere placed in a microwave cavity near the maximum magnetic field  $\zeta$  for the cavity mode and simultaneously in a uniform bias magnetic field establishing the magnon photon phonon coupling and The cavity system is probed through a probe input on one side $\eta_{p}$ with a traveling field at angle $\theta$ along $x$-axis and $y$-axis, respectively.}
	\label{fig1}
\end{figure}

We consider a single-crystal yttrium iron garnet (YIG) microsphere with a diameter ranging from $10^{2}\mathrm{\mu m}$ to $1\mathrm{mm}$, positioned inside a Fabry–Pérot cavity of length $L\approx12.5\times10^{-4}\mathrm{m}$, as illustrated in Fig.~\ref{fig1}~\cite{Lai2024}. A bias magnetic field of about $10,\mathrm{G}$ is applied perpendicular to the cavity axis to resonantly drive the YIG sphere, exciting collective spin precession (magnons) described by the operator $\hat{m}$. The coupling between $B_{0}$ and $\hat{m}$ is given by $\eta=(\sqrt{5}/4)\gamma_{B}\sqrt{N}B_{0}$, where the system is driven with power $P_{0}\approx0.0164\mathrm{mW}$ at frequency $\omega_{0}\approx3.8\times2\pi\times10^{8}\mathrm{Hz}$, and the gyromagnetic ratio is $\gamma_{B}/2\pi=28\mathrm{GHz/T}$. The total spin number is $N=\rho V$, with spin density $\rho=2\times10^{22}\mathrm{cm^{-3}}$ and sample volume $V$. The uniform magnon mode oscillates at $\omega_{m}\approx2\pi\times2\mathrm{MHz}$ with a decay rate $\kappa_{m}\approx2\pi\times3\mathrm{MHz}$. The magnons interact with the pre-existing microwave cavity photons through magnetic dipole coupling, yielding the magnon–photon coupling rate $g_{a}=(g_{s}\mu_{B}/2\hbar)\sqrt{\mu_{0}\hbar\omega_{c}V_{s}/V}$, where $g_{s}$ is the electron $g$-factor and $\mu_{B}$ is the Bohr magneton. For $V/V_{s}\approx1$, this coupling reaches values on the order of $10,\mathrm{MHz}$, corresponding to the strong-coupling regime. Moreover, the magnetostrictive interaction between the magnetic field and the spin ensemble excites a mechanical vibration (phonon mode $\hat{b}$) in the YIG, resulting in a magnon–phonon coupling $G_{b}\propto1/D$, inversely proportional to the YIG diameter~\cite{Zhang2016_SciAdv}. For $D\approx250\mathrm{\mu m}$, the magnon-phonon coupling will be $G_{b}\approx2\pi\times9.88\mathrm{mHz}$, with mechanical frequency $\omega_{b}\approx2\pi\times9.88\mathrm{MHz}$, and dissipation $\gamma_{a}\approx2\pi\times300\mathrm{Hz}$. The magnomechanical coupling can be further enhanced by reducing the sphere’s size, increasing both strain sensitivity and phonon participation. In our analysis, we normalize the magnon–phonon coupling by the magnon detuning, $G_{b}/\Delta_{m}$. This choice highlights that the effective strength of the magnomechanical interaction can be enhanced experimentally by reducing the effective detuning $\Delta_{m}$, for example via magnetic-field tuning.

A coherent microwave source also interacts with the system, exciting cavity photons $\hat{a}$ whose wavelength is much larger than the YIG diameter, ensuring negligible radiation-pressure effects. In addition to this drive, a weak probe microwave field $\eta_{p}$ is applied (see Fig.~\ref{fig1}) to investigate two-photon excitation processes and magnon-mediated interference phenomena~\cite{Yasir2022, Yasir2016, Yasir2017, Yasir2023}. When the probe frequency approaches the hybrid magnon–phonon resonance, two excitation pathways—direct cavity excitation and indirect magnon–phonon coupling—interfere destructively, producing a narrow transparency window in the cavity transmission spectrum. This interference mechanism, analogous to electromagnetically induced transparency (EIT) in atomic systems, enables controllable suppression of probe absorption and reveals coherent information exchange between the cavity, magnon, and phonon subsystems. All parameters used here are consistent with well-established experimental conditions~\cite{Zhang2016_SciAdv, Weis2010}. The total Hamiltonian of the system reads as,

 \begin{eqnarray}
  &&H_{\text{total}} = H_0 + H_{\text{int}} + H_d + H_{\text{p}}, \\
  &&H_0  = \omega_{ a} \hat{a} ^{\dagger}  \hat{a} +\omega _{m}  \hat{m} ^{\dagger} \hat{m}+ \frac{\omega_{b}}{2}  (\hat{p} ^2 + \hat{q} ^2 ) \\
  &&H_{\text{int}}  = g_{b} \hat{m} ^{\dagger} \hat{mq} + G_{a} (\hat{a}+ \hat{a}^ \dagger )(\hat{m}+ \hat{m} ^\dagger )\nonumber\\
  &&-i\Gamma e ^{i\theta } (\hat{a} ^\dagger \hat{m} + \hat{m} ^\dagger \hat{a})\\
  &&H_{\text{drive}} = i\eta (\hat{m} ^\dagger e ^{-i\omega {_o} t} - \hat{m}e ^{i\omega {_o}t})\\
  &&H_{\text{probe}}  = i\eta  _{p} (\hat{a}^\dagger e ^{-i\Delta  _{p} t}- \hat{a}e ^{i\Delta  _{p} t})       
  \end{eqnarray}
Where $\hat{a},(\hat{a}^\dagger)$, $\hat{m},(\hat{m}^\dagger)$, and $\hat{b},(\hat{b}^\dagger)$ denote the annihilation (creation) operators of the cavity, magnon, and phonon modes, respectively, with the standard commutation relation $[\hat{O},\hat{O}^\dagger]=1$ for $O=a,m,b$. The parameter $\Gamma=\alpha\sqrt{\hbar/(\omega_{m}m_{m})}$ represents the coupling strength between the traveling field and the magnon mode, where $\alpha$ is the amplitude of the traveling field. The quantities $\delta$ and $\theta$ correspond to the frequency detuning and the incidence angle of the traveling field relative to the cavity axis.

By applying the rotating-wave approximation, in which $(\hat{a}+\hat{a}^\dagger)(\hat{m}+\hat{m}^\dagger)\rightarrow(\hat{a}^\dagger\hat{m}+\hat{a}\hat{m}^\dagger)$, and assuming the phase factor $e^{\delta t+i\theta}$ to be time-independent ($e^{\delta t+i\theta}!\rightarrow!e^{i\theta}$), the total Hamiltonian can be rewritten in a simplified form suitable for describing the resonant magnon–photon interactions and the interference processes leading to electromagnetically induced transparency.
 \begin{eqnarray}
  H_{\text{total}} &=& \Delta_{a} \hat{a} ^{\dagger}  \hat{a} +\Delta _{m}  \hat{m} ^{\dagger} \hat{m}+ \frac{\omega_{b}}{2}  (\hat{p} ^2 + \hat{q} ^2 )+ g_{b} \hat{m} ^{\dagger} \hat{mq}\nonumber\\
   &&+ G _{a} (\hat{a}+ \hat{a}^ \dagger )(\hat{m}+ \hat{m} ^\dagger ) -i\Gamma e ^{i\theta } (\hat{a} ^\dagger \hat{m} + \hat{m} ^\dagger \hat{a})\nonumber\\
 &&+i\eta  _{p} (\hat{a}^\dagger e ^{-i\Delta  _{p} t}- \hat{a}e ^{i\Delta  _{p} t}) \nonumber\\
 &&+i\eta (\hat{m} ^\dagger e ^{-i\omega {_o} t} - \hat{m}e ^{i\omega {_o}t}),
\end{eqnarray}
where $\Delta_{a} = \omega_{a} - \omega_{0}$ and
$\Delta_{m} = \omega_{m} - \omega_{0}$ denote the detunings
of the cavity and magnon modes from the drive frequency $\omega_{0}$, respectively. In order to govern the integrated hybrid response of the system, we compute the quantum Language equation from the total Hamiltonian,
\begin{eqnarray}
\dot{\hat{a}} & = & -(i\Delta_a+\kappa_a )\hat{a} -(iG _{a} +\Gamma e ^{i\theta })\hat{m}+ \eta  _{p} e ^{-i\Delta  _{p} t} + \sqrt{2\kappa_{a }}a^{in},\nonumber\\
\dot{\hat{m}} & = & -(i\Delta_m+\kappa_m - g _{b} {q})\hat{m} -(iG _{a}  +\Gamma e ^{i\theta })\hat{a} + \eta + \sqrt{2\kappa_{m }}m^{in},\nonumber\\
\dot{\hat{p}} & = & -\omega _{b}\hat{q}- g _{b} {\hat{m}} ^{\dagger}{\hat{m}}-\gamma _{b}\hat{p},\nonumber\\
\dot{\hat{q}} & = & \omega _{b}\hat{p}.
\end{eqnarray}
Here, $\kappa_{a}$, $\kappa_{m}$, and $\gamma_{b}$ denote the dissipation rates of the cavity, magnon, and mechanical modes, respectively. The operators $\hat{a}^{\mathrm{in}}$, $\hat{m}^{\mathrm{in}}$, and $\hat{\xi}$ represent the corresponding input noise operators, which are modeled as white Gaussian noise fields with zero mean and standard correlation relations
$\langle \hat{O}^{\dagger}{\mathrm{in}}(t)\hat{O}{\mathrm{in}}(t') + \hat{O}{\mathrm{in}}(t')\hat{O}^{\dagger}{\mathrm{in}}(t) \rangle = (2\bar{n}{o}+1)\delta(t-t')$,
where $o={a,m,b}$ and $\bar{n}{o} = [\exp(\hbar\omega_{o}/k_{B}T)-1]^{-1}$ is the mean thermal occupation of the corresponding mode~\cite{Yasir2022, Lai2024, Aspelmeyer2014}.

To analyze the system dynamics near the steady state, we linearize the quantum Langevin equations by expressing each operator as $\hat{O}=O_{s}+\delta\hat{O}(t)$, where $O_{s}$ is the steady-state mean value and $\delta\hat{O}(t)$ represents the first-order quantum fluctuation. Second-order fluctuation terms are neglected as they are typically much smaller in magnitude and do not contribute significantly to linear response phenomena such as electromagnetically induced transparency (EIT). Consequently, the linearized quantum Langevin equations governing the quadrature fluctuations $(\delta\hat{a}, \delta\hat{a}^{\dagger}, \delta\hat{m}, \delta\hat{m}^{\dagger}, \delta\hat{q}, \delta\hat{p})$ can be written as follows.
\begin{eqnarray}
\delta \dot{\hat{a}} & = & -(i\Delta_a+\kappa_a )\delta \hat{a} -(iG _{a}  +\Gamma e ^{i\theta })\delta \hat{m}\\
\delta\dot{\hat{m}} & = & -(i\Delta_m+\kappa_m)\delta \hat{m}- iG_{mb} \delta{\hat{q}} -(iG _{a}  +\Gamma e ^{i\theta })\delta \hat{a}\\
\delta\dot{\hat{p}} & = & -\omega _{b}\delta \hat{q}+G_{b}( \delta{\hat{m}} +\delta{\hat{m}}^{\dagger })-\gamma _{b}\delta \hat{p} \\
\delta\dot{\hat{q}} & = & \omega _{b}\delta \hat{p}
\end{eqnarray}
where $ G_{b}=g_{mb}\langle m \rangle $ and $ \Delta_{a,m,b}=\omega_{a,m,b}-\omega_0$. This treatment effectively isolates the coherent response of the system from stochastic noise processes. Since the EIT effect arises primarily from interference between coherent excitation pathways, and the noise terms average to zero in the mean-field limit, neglecting them provides an accurate and experimentally justified approximation under typical low-temperature, weak-noise conditions. One can approximate, The effective Hamiltonian matrix $H_{eff}$ from above linearized quantum Langevin equations reading as,
\begin{equation}
\label{pt}
H_\text{eff} =
\begin{pmatrix}
	\Delta_{a} + i\kappa_a & G_a + i\Gamma e^{i\theta} & 0\\
	G_a + i\Gamma e^{i\theta} & \Delta_{m} + i\kappa_m & G_b\\
	0 & G_{mb} & \omega_b + i\gamma_b
\end{pmatrix}.
 \end{equation}
We omit vector notation for the bosonic field operators for simplicity.
We introduce the parity operator $\mathcal{P}$ such that it exchanges the
cavity and magnon modes according to
$\mathcal{P}:\hat{a} \leftrightarrow -\hat{m}$ and
$\hat{a}^\dagger \leftrightarrow -\hat{m}^\dagger$.
The time-reversal operator $\mathcal{T}$ acts as
$\mathcal{T}: i \rightarrow -i$ while leaving $\hat{a}$ and $\hat{a}^\dagger$
invariant and reversing the sign of the magnon quadratures,
$\hat{m} \rightarrow -\hat{m}$ and $\hat{m}^\dagger \rightarrow -\hat{m}^\dagger$.
Using these transformations, one verifies that
$\hat{H}_\text{eff}^{\mathcal{PT}} = (\mathcal{PT}) \hat{H}_\text{eff} (\mathcal{PT})^{-1}
= \hat{H}_\text{eff}$, confirming the effective $\mathcal{PT}$ symmetry of the
Hamiltonian in Eq.~(12)~\cite{Peng2014,Ozdemir2019,Xu2021}.
\begin{figure*}[tp]
	\includegraphics[width=16cm]{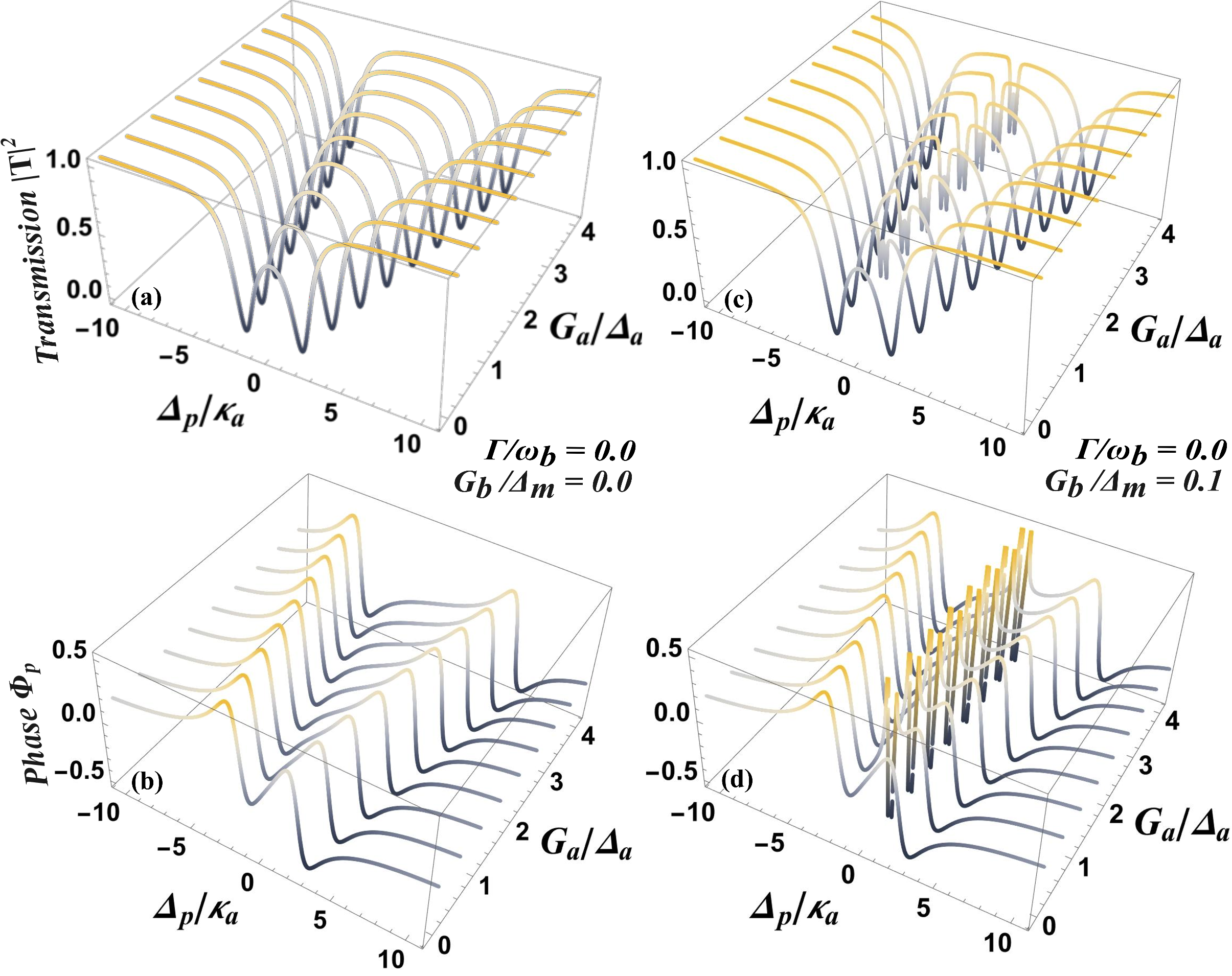}
	\caption{(Color online) Normalized cavity probe transmission 
		$T(\Delta_{p})$ and phase $\phi(\Delta_{p})$ in the Hermitian regime ($\Gamma=0$). 
		(a),(b) Single-window MMIT for $G_{b}/\Delta_{m}=0$ and increasing photon--magnon coupling 
		$G_{a}/\Delta_{a}$ (values indicated). 
		(c),(d) Double-window MMIT for finite magnomechanical coupling $G_{b}/\Delta_{m}\neq 0$, with 
		$G_{a}/\Delta_{a}$ unchanged. Introducing $G_{b}$ splits the single transparency resonance into 
		two magnomechanical transparency windows and gives rise to two corresponding steep dispersive 
		features in the phase. The other numerical parameters that we consider in our calculations are  $\kappa_a/\Delta_a=0.08,\kappa_m/\Delta_m=0.08$} 
	\label{Fig2}
\end{figure*}

To analyze the quantum interference that gives rise to electromagnetically induced transparency (EIT), we evaluate the cavity transmission spectrum. Under the condition $\eta \gg \eta_{p}$, the linearized quadrature fluctuations of each subsystem are expanded as $\delta\mathcal{B}(t)=\sum_{n=\{+,-\}}\mathcal{B}_{n}e^{in\omega t}$, where $\mathcal{B}$ denotes a generic operator corresponding to the cavity, magnon, or mechanical mode. Substituting this expansion into the linearized Langevin equations and comparing the coefficients associated with the probe-dependent exponential terms allows us to extract the system response at the probe frequency. The resulting coefficient multiplying the $e^{-i\omega t}$ component, denoted as $c_{-}$, encapsulates the contribution of coherent interference between the cavity, magnonic, and mechanical excitation pathways. This term therefore governs the appearance of the EIT window and determines how the probe field experiences transparency, amplification, or dispersion within the non-Hermitian cavity magnomechanical system.
\begin{equation}
	c_{-}(\Delta_{p})=\frac{\eta_{p}\,\mathcal{X}_{m}(\Delta_{p})}
	{\bigl(\kappa_{a}+i(\Delta_{a}-\Delta_{p})\bigr)\,\mathcal{X}_{m}(\Delta_{p})
		- \bigl(iG_{a}+\Gamma e^{i\theta}\bigr)^{2}},
	\label{cminus-new}
\end{equation}
where,
\begin{eqnarray}
	\mathcal{X}_{m}(\Delta_{p})&=&\kappa_{m}+i(\Delta_{m}-\Delta_{p})
	+ \frac{i\,G_{mb}G_{b}\,\omega_{b}}{\mathcal{X}_{b}(\Delta_{p})},\\
	\mathcal{X}_{b}(\Delta_{p})&=&\omega_{b}^{2}-\Delta_{p}^{2}-i\gamma_{b}\Delta_{p}.
\end{eqnarray}
Here $\mathcal{X}_{m}(\Delta_p)$ and $\mathcal{X}_{b}(\Delta_p)$ defines the effective susceptibilities for magnons and phonons interacting the probe field. 

Furthermore, to evaluate the probe-field contribution to the cavity transmission, we make use of the standard input--output relation $c_{\mathrm{out}}=\sqrt{2\kappa}\,c-c_{\mathrm{in}}$, which gives
\begin{eqnarray}
	\mathcal{T}(\omega_{p})
	&=&\frac{\eta_{p}-\sqrt{2\kappa}\,c_{-}(\Delta_{p})}{\eta_{p}}
	\nonumber\\
	&=&1-\frac{\sqrt{2\kappa}\,c_{-}(\Delta_{p})}{\eta_{p}},
\end{eqnarray}
with transmission phase $\Phi_p(\omega_p)=arg(\mathcal{T})$. The corresponding transmission amplitude,
\begin{equation}
	\mathcal{E}_{\mathrm{out}}(\Delta_{p})
	=\frac{\sqrt{2\kappa}\,c_{-}(\Delta_{p})}{\eta_{p}},
\end{equation}
encodes the full probe response of the cavity. Its real part describes the absorptive (in-phase) component, whereas its imaginary part represents the dispersive (out-of-phase) behavior. In this form, $\mathcal{E}_{\mathrm{out}}(\Delta_{p})$ captures the characteristic modifications of transparency, absorption, and phase arising from the coupled cavity--magnon--phonon dynamics. If both the magnon and mechanical modes are decoupled from the cavity, the probe transmission reduces to,
\begin{equation}
	\mathcal{T}(\Delta_p)=\frac{\sqrt{2\kappa}}{\kappa + i(\Delta - \Delta_p)}.
\end{equation}

\section{Non-Hermitian Magnomechanically induced transparencies}\label{sec2}
MMIT provides a direct means of probing coherent interference in hybrid magnonic platforms and plays a central role in understanding how cavity photons, magnons, and mechanical vibrations interact in the linearized regime. In cavity magnomechanics, MMIT emerges from destructive interference between a cavity--magnon excitation pathway and a long-lived magnonic (or magnomechanical) coherence, closely paralleling the mechanism of optomechanically induced transparency in cavity optomechanics. The ability to generate, tune, and split MMIT windows is therefore essential for applications in slow light, microwave photonic signal processing, and quantum transduction. Importantly, the present system, even in the absence of engineered non-Hermiticity ($\Gamma=0$), provides a robust route to realizing MMIT, because the cavity--magnon coupling $G_{a}$ can be made large, and the mechanical mode can be hybridized with magnons through the magnetostrictive interaction.

Figures~\ref{Fig2}(a)--\ref{Fig2}(d) illustrate these interference processes by showing the normalized probe transmission $T(\Delta_{p})$ and the corresponding phase response $\phi(\Delta_{p})$ for different choices of the photon--magnon and magnomechanical couplings. Figures~\ref{Fig2}(a) and~\ref{Fig2}(b) show the Hermitian baseline case with $G_{b}/\Delta_{m}=0$ and $\Gamma=0$. For small $G_{a}/\Delta_{a}$, the cavity spectrum remains essentially Lorentzian. As $G_{a}/\Delta_{a}$ increases, a sharp transparency window forms around $\Delta_{p}\approx 0$, characteristic of a single MMIT window. The transparency depth and the associated dispersive slope increase monotonically with $G_{a}/\Delta_{a}$, indicating the strengthening of coherent interference between the cavity and magnon channels. The phase profile in Fig.~\ref{Fig2}(b) displays the corresponding steep dispersive feature, which is a hallmark of EIT-like processes and reflects a large group delay experienced by the probe within the transparency window.
\begin{figure}[tp]
	\includegraphics[width=8.5cm]{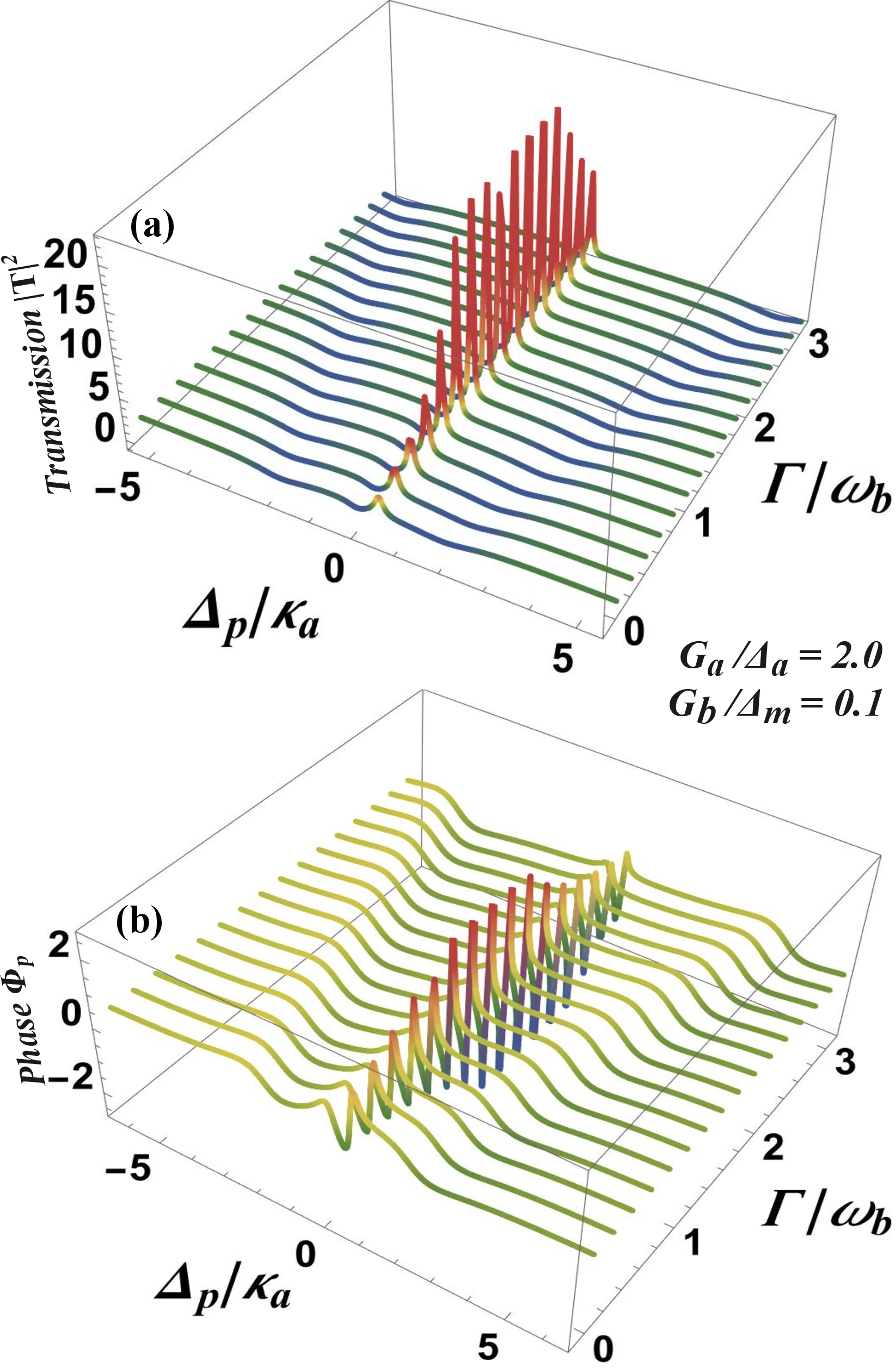}
	\caption{(Color online) Normalized cavity probe (a) transmission $T(\Delta_{p})$ and (b) phase $\phi(\Delta_{p})$ as functions of the normalized probe detuning $\Delta_{p}/\omega_{b}$ for nonzero non-Hermitian coupling $\Gamma$, with $G_{a}/\Delta_{a}=2$ and $G_{b}/\Delta_{m}=0.1$. The presence of $\Gamma\neq 0$ converts the Hermitian MMIT window into a gain-assisted transparency feature with $T(\Delta_{p})>1$ and a strongly asymmetric line shape: transmission is suppressed on the loss-dominated (negative-detuning) side and amplified on the gain-dominated (positive-detuning) side, accompanied by a sharply varying phase dispersion near the amplified resonance. The remaining parameters are the same as in Fig.~\ref{Fig2}.}
	\label{Fig3}
\end{figure}

The situation changes qualitatively when the magnomechanical interaction is introduced. In Figs.~\ref{Fig2}(c) and~\ref{Fig2}(d), the magnomechanical coupling is set to a finite value $G_{b}/\Delta_{m}\neq 0$ while the system remains Hermitian ($\Gamma=0$). Because the magnon mode is now hybridized with the mechanical mode, the effective magnonic resonance splits into two normal-mode branches. Consequently, the probe field interacts with two distinct magnon--phonon pathways, and the single MMIT peak evolves into a two-window transparency structure. Figures~\ref{Fig2}(c) and~\ref{Fig2}(d) clearly reveal this doublet: two resolved transparency windows appear at separated detunings, with an absorption region between them. Each transparency window is accompanied by its own steep phase dispersion, as seen in Fig.~\ref{Fig2}(d). This double-window behavior is analogous to a magnomechanical Autler--Townes splitting and demonstrates that the mechanical mode can be used as an independent and highly tunable control knob for shaping MMIT.

It is worth emphasizing that, in standard cavity-magnomechanical experiments, the bare magnomechanical coupling $G_{b}$ is typically much smaller than the photon--magnon coupling $G_{a}$, i.e., $G_{b}\ll G_{a}$. In our analysis, we employ \emph{scaled} values of $G_{b}/\Delta_{m}$ to make the double-window structure clearly visible at the level of theory. Such enhanced effective magnomechanical coupling can be approached experimentally by reducing the YIG sphere diameter, improving the mechanical quality factor, or enhancing the spatial overlap between magnon and phonon modes. Therefore, the double-window MMIT predicted here is not only conceptually important but also experimentally accessible with current cavity-magnomechanics technology.

In the non-Hermitian cavity magnomechanical regime, MMIT provides a natural platform for realizing gain-assisted transparency and controllable amplification. When the traveling-field–induced non-Hermitian coupling $\Gamma$ is nonzero, the effective magnon--photon dimer acquires an imbalance of gain and loss that can be tuned across and beyond the $\mathcal{PT}$-symmetry-breaking threshold identified in Ref.~\cite{Chengyong2025}. In this regime, the MMIT window is no longer a purely passive interference feature: the same coherent pathway that produces transparency in the Hermitian limit can be converted into an amplifying channel, such that the probe experiences transmission $T(\Delta_{p})>1$ around selected detunings. The small but finite magnomechanical coupling $G_{b}/\Delta_{m}=0.1$ considered here does not yet split the transparency into multiple windows, but it does provide an additional feedback route that slightly reshapes the line profile and enhances the sensitivity of the response to the non-Hermitian parameters.

Figure~\ref{Fig3} shows the normalized cavity probe transmission $T(\Delta_{p})$ and phase $\phi(\Delta_{p})$ as functions of the normalized probe detuning $\Delta_{p}/\omega_{b}$ for nonzero non-Hermitian coupling, with $G_{a}/\Delta_{a}=2$ and $G_{b}/\Delta_{m}=0.1$ fixed. In Fig.~\ref{Fig3}(a), one observes that, unlike the purely Hermitian case, the MMIT feature is accompanied by a clear amplification of the probe signal: near the central hybrid resonance, the transmission peak exceeds unity, reflecting the net gain introduced by the traveling field in the coupled magnon--photon subsystem. Spectrally, the line shape becomes asymmetric with respect to $\Delta_{p}=0$. On the negative-detuning side, where the effective mode is predominantly loss-dominated, the transmission is suppressed and the MMIT dip is broadened, indicating stronger effective dissipation. On the positive-detuning side, by contrast, the gain-dominated hybrid mode is preferentially excited; interference between the direct cavity pathway and the magnon--mechanical channel then produces a pronounced, narrowed transparency peak with $T(\Delta_{p})>1$, characteristic of gain-assisted MMIT in a $\mathcal{PT}$-broken regime.

The corresponding phase response in Fig.~\ref{Fig3}(b) exhibits a markedly non-Lorentzian dispersion. Around the amplified transparency region on the positive-detuning side, the phase shows a sharp and highly localized variation, signaling a strongly enhanced group delay and a high sensitivity of the probe phase to small changes in detuning. This steep phase slope is consistent with interference near an exceptional-point–like configuration, where the hybrid eigenmodes become nearly coalescent in both frequency and linewidth. On the negative-detuning side, where absorption dominates, the phase varies more smoothly, reflecting the predominance of a lossy hybrid mode. Taken together, Figs.~\ref{Fig3}(a) and~\ref{Fig3}(b) demonstrate that introducing a non-Hermitian coupling $\Gamma$ transforms conventional MMIT into a non-Hermitian, gain-assisted transparency with strong spectral asymmetry and controllable amplification, while a small but finite $G_{b}/\Delta_{m}$ ensures that the magnomechanical backaction remains present but does not obscure the essential non-Hermitian features.
\begin{figure}[tp]
	\includegraphics[width=8.5cm]{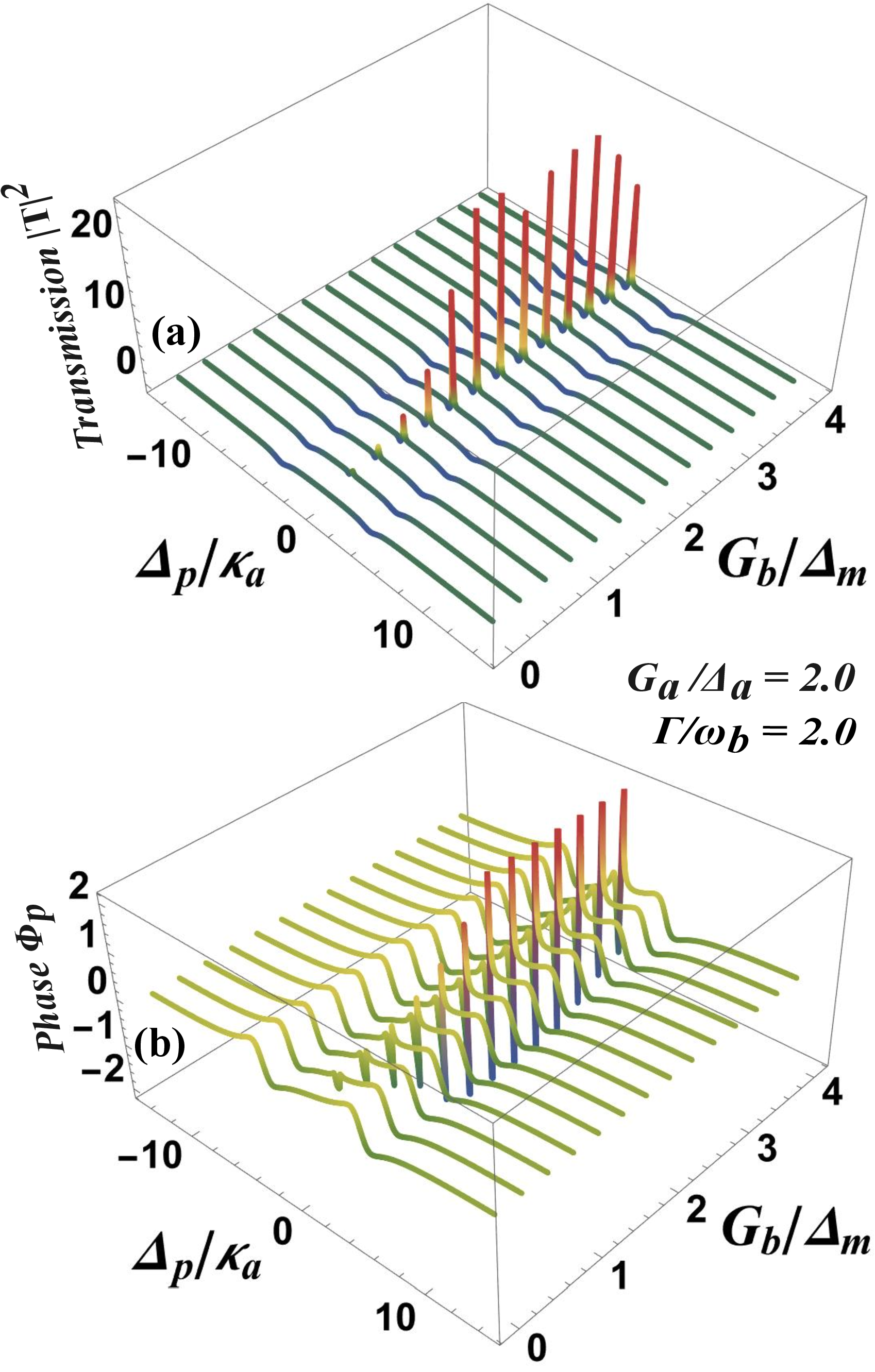}
	\caption{(Color online) Cavity probe transmission $T(\Delta_{p})$ and phase $\phi(\Delta_{p})$ as functions of the probe detuning $\Delta_{p}$ for fixed non-Hermitian parameter $\Gamma/\omega_{b}=2$ and photon--magnon coupling $G_{a}/\Delta_{a}=2$, and for different magnomechanical couplings $G_{b}/\Delta_{m}$ (as indicated in the legend). Increasing $G_{b}/\Delta_{m}$ enhances the magnomechanical backaction on the non-Hermitian hybrid modes, leading to a pronounced reshaping of the gain-assisted MMIT profile and a steeper, more asymmetric phase dispersion near the amplified transparency region. The remaining parameters are the same as in Fig.~\ref{Fig2}.}
	\label{Fig4}
\end{figure}
We now investigate how the magnomechanical coupling modifies non-Hermitian MMIT and the associated gain properties of the probe field. In the effective description, the mechanical mode couples to the magnon via $G_{b}$ and enters the magnon dynamics as a complex self-energy, renormalizing both the resonance frequency and the effective linewidth of the hybrid magnon--photon subsystem. When the non-Hermitian parameter $\Gamma$ is finite, this mechanical self-energy competes with and reshapes the gain--loss imbalance generated by the traveling field. As a consequence, tuning $G_{b}/\Delta_{m}$ at fixed $\Gamma/\omega_{b}$ and $G_{a}/\Delta_{a}$ provides a convenient way to control the spectral profile of non-Hermitian MMIT, including the position, width, and magnitude of the amplification peak. We reiterate that, in typical cavity-magnomechanics experiments, the bare magnomechanical coupling is much weaker than the photon--magnon coupling ($G_{b}\ll G_{a}$). In our analysis, however, we employ scaled values of $G_{b}/\Delta_{m}$ to clearly resolve its impact on the gain-assisted transparency. Such regimes can be approached experimentally by reducing the YIG size, increasing the mechanical quality factor, and optimizing the spatial overlap between magnon and phonon modes.
\begin{figure}[tp]
	\includegraphics[width=8.5cm]{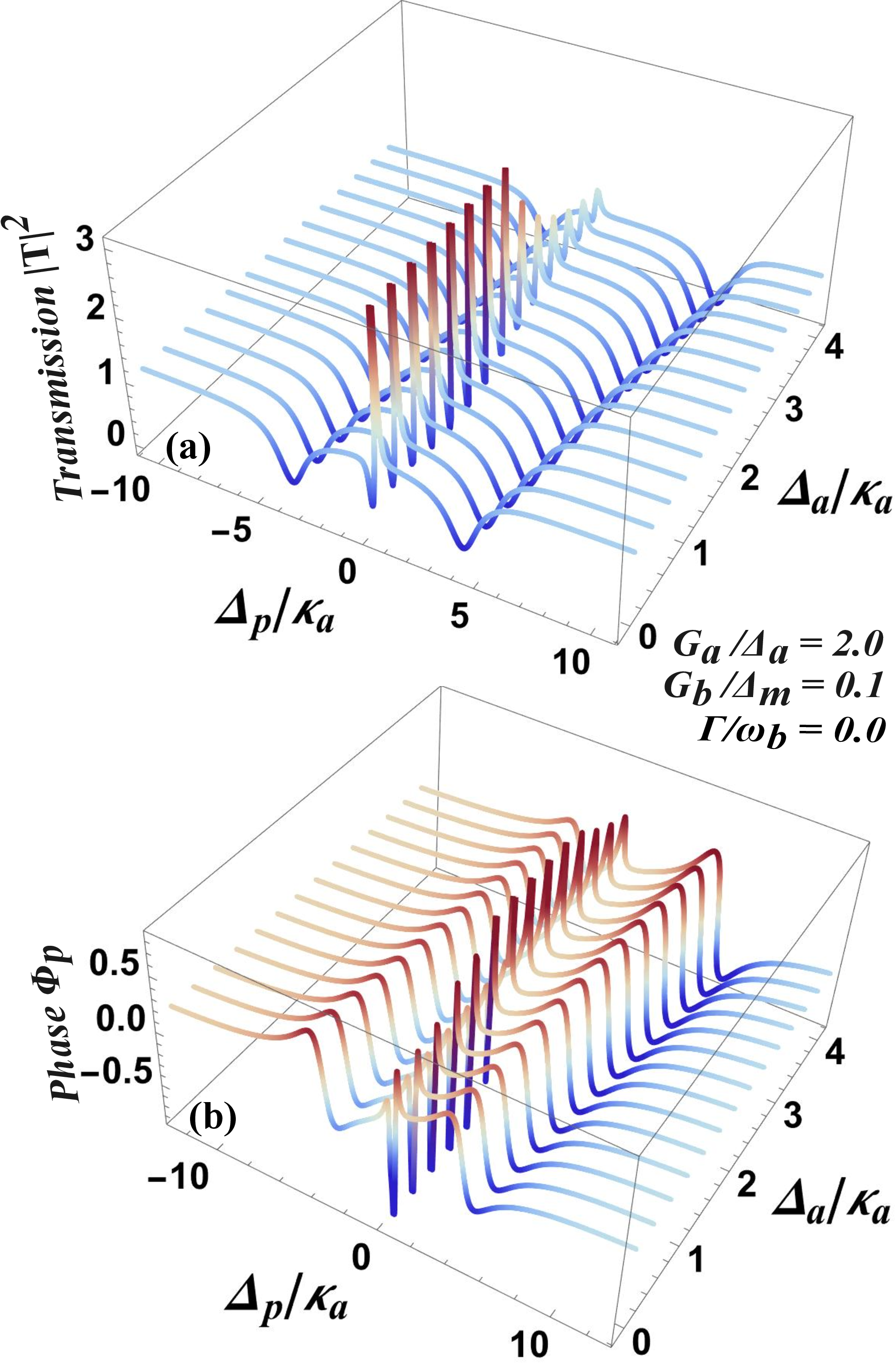}
	\caption{(Color online) (a) Cavity probe transmission $T(\Delta_{p})$ and (b) phase response $\phi(\Delta_{p})$ for different cavity detunings $\Delta_{a}/\kappa_{a}$ (values indicated in the legend), at fixed non-Hermitian coupling $\Gamma/\kappa_{a}=2$, photon--magnon coupling $G_{a}/\kappa_{a}=2$, and magnomechanical coupling $G_{b}/\kappa_{a}=0.1$. Varying $\Delta_{a}/\kappa_{a}$ transforms the MMIT-like feature into asymmetric Fano profiles and produces strongly non-Lorentzian phase dispersion, with amplification on one side of the resonance and enhanced absorption on the other. The remaining parameters are the same as in Fig.~\ref{Fig2}.}
	\label{Fig5}
\end{figure}

Figure~\ref{Fig4} displays the normalized cavity probe transmission $T(\Delta_{p})$ and phase $\phi(\Delta_{p})$ as functions of the probe detuning $\Delta_{p}$ for a fixed non-Hermitian strength $\Gamma/\omega_{b}=2$ and photon--magnon coupling $G_{a}/\Delta_{a}=2$, while the magnomechanical coupling $G_{b}/\Delta_{m}$ is varied as indicated in the legend. In Fig.~\ref{Fig4}(a), for smaller $G_{b}/\Delta_{m}$ the transmission exhibits a single, strongly asymmetric MMIT-like feature with a clear amplification region near the hybrid resonance. This amplified transparency arises from constructive interference between the probe-driven cavity field and the gain-enhanced magnon channel in the $\mathcal{PT}$-broken regime, in a situation where the magnon--phonon backaction only weakly perturbs the effective non-Hermitian dimer formed by the cavity and magnon modes. The corresponding phase response in Fig.~\ref{Fig4}(b) shows a steep, non-Lorentzian dispersion localized around the amplified region, indicating a large and tunable group delay (or advance) for the transmitted probe.
\begin{figure*}[tp]
	\includegraphics[width=17cm]{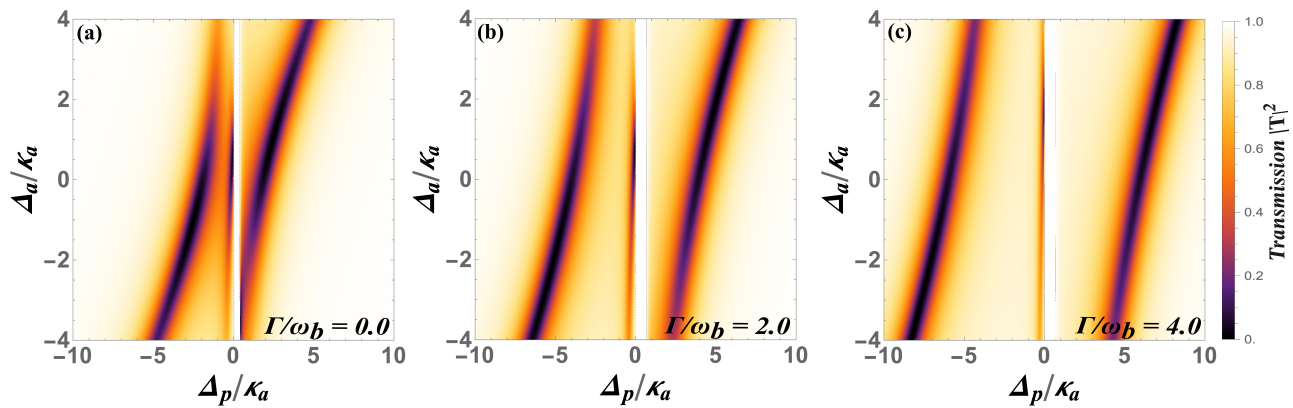}
	\caption{(Color online) Density plots of the cavity probe transmission $T(\Delta_{p},\Delta_{m})$ as functions of probe detuning $\Delta_{p}/\kappa_{a}$ and magnon detuning $\Delta_{m}/\kappa_{a}$ for (a) $\Gamma=0$, (b) $\Gamma/\omega_{b}=2$, and (c) $\Gamma/\omega_{b}=4$. The photon--magnon and magnomechanical couplings are fixed at $G_{a}/\Delta_{a}=2$ and $G_{b}/\Delta_{m}=0.1$, respectively. Increasing $\Gamma$ deforms the MMIT/Fano structures from nearly symmetric interference patterns into strongly asymmetric, gain-assisted Fano ridges with enhanced contrast between high- and low-transmission regions. The remaining parameters are the same as in Fig.~\ref{Fig2}.}
	\label{Fig6}
\end{figure*}

As $G_{b}/\Delta_{m}$ is increased, the influence of the mechanical mode on the non-Hermitian hybrid dynamics becomes more pronounced. The enhanced magnomechanical self-energy significantly modifies the effective magnon susceptibility, redistributing gain and loss among the hybridized modes. In the transmission spectrum, this manifests as a clear reshaping of the amplification profile: the gain peak becomes spectrally narrower and can shift in detuning, and additional structure may emerge in the form of shoulders or partial splitting around the central transparency region. Physically, this reflects the fact that the probe field now interferes with a more strongly dressed magnon--phonon pathway, so that the balance between destructive and constructive interference becomes strongly detuning dependent. On the side of the spectrum where the gain-dominated hybrid branch prevails (typically at positive detuning), amplification is enhanced and more sharply localized, whereas on the opposite side the increased coupling to the mechanical mode can reinforce effective loss and reduce the transmission. The phase response tracks these changes by developing sharper and more asymmetric variations: for larger $G_{b}/\Delta_{m}$, the phase jump becomes both steeper and compressed in frequency, signaling an increased sensitivity of the probe phase to small changes in detuning. Altogether, these results demonstrate that, at fixed $\Gamma/\omega_{b}$ and $G_{a}/\Delta_{a}$, the magnomechanical coupling $G_{b}/\Delta_{m}$ acts as an additional control knob to tailor the strength and bandwidth of non-Hermitian MMIT amplification and to engineer highly dispersive transmission windows suitable for tunable quantum amplification and slow-/fast-light control in cavity magnomechanical systems.

\section{Fano resonances}\label{sec3}
Fano resonances arise from quantum interference between a narrow discrete resonance and a broad continuum (or quasi-continuum) background, leading to characteristically asymmetric line shapes rather than simple Lorentzian peaks or dips~\cite{Ref280,Ref2800}. In hybrid light--matter systems, such as optomechanical or magnomechanical platforms, Fano profiles typically emerge when a sharply defined mode (e.g., a mechanical or magnonic resonance) interferes with a broader cavity response. The resulting spectral asymmetry is highly sensitive to the relative phase and amplitude of the interfering pathways and can be tuned by external control parameters such as detunings, coupling strengths, or non-Hermitian gain--loss terms. This sensitivity makes Fano resonances particularly attractive for precision spectroscopy, switchable transparency, and enhanced sensing, especially in regimes where non-Hermitian effects further distort and amplify the interference structure.

In our non-Hermitian cavity magnomechanical system, the Fano behavior is governed by the interference between the cavity--magnon channel and the weakly coupled mechanical mode in the presence of a finite gain--loss parameter $\Gamma$. Figure~\ref{Fig5} illustrates this behavior for fixed non-Hermitian coupling $\Gamma/\kappa_{a}=2$, photon--magnon coupling $G_{a}/\kappa_{a}=2$, and relatively weak magnomechanical coupling $G_{b}/\kappa_{a}=0.1$, while the normalized cavity detuning $\Delta_{a}/\kappa_{a}$ is varied. As discussed earlier, a finite $\Gamma$ drives the system into a non-Hermitian regime where the hybrid eigenmodes acquire unequal effective linewidths, so that the interference between the narrow magnomechanical pathway and the broader cavity background naturally gives rise to Fano-type line shapes. In Fig.~\ref{Fig5}(a), the cavity probe transmission $T(\Delta_{p})$ evolves from a nearly symmetric MMIT-like feature to distinctly asymmetric Fano profiles as $\Delta_{a}/\kappa_{a}$ is tuned away from zero. For certain values of $\Delta_{a}/\kappa_{a}$, one side of the transparency feature is strongly enhanced while the other side becomes more absorptive, yielding the characteristic peak--dip or dip--peak structure of a Fano resonance. The asymmetry reverses when the sign of $\Delta_{a}/\kappa_{a}$ is changed, reflecting the change in relative phase between the cavity-like and magnon--mechanical branches. The presence of nonzero $\Gamma$ further sharpens and distorts these profiles, so that on one side of the resonance the interference can even lead to amplification of the transmitted probe, whereas on the opposite side it reinforces absorption.

The corresponding phase response in Fig.~\ref{Fig5}(b) provides complementary insight: instead of a single, smooth phase jump associated with a symmetric MMIT feature, the phase now exhibits a strongly non-Lorentzian, multi-step variation correlated with the Fano distortion in the transmission. Around the frequency where destructive interference is strongest, the phase displays a rapid and pronounced change, indicative of a large and rapidly varying group delay. On the opposite side, where constructive interference dominates and amplification is observed, the phase also varies steeply but with opposite curvature, consistent with the characteristic dispersive structure of Fano resonances. Together, Figs.~\ref{Fig5}(a) and~\ref{Fig5}(b) demonstrate that tuning the cavity detuning $\Delta_{a}/\kappa_{a}$ in the presence of non-Hermitian coupling and weak magnomechanical interaction enables controlled sculpting of asymmetric Fano line shapes and their associated phase dispersion, providing an additional handle for engineering highly sensitive, interference-based functionalities in cavity magnomechanics.

To obtain a global picture of how Fano-type interference develops and is reshaped by non-Hermiticity in our hybrid cavity--magnon--phonon system, we plot in Fig.~\ref{Fig6} the density maps of the cavity probe transmission as a function of both the probe detuning $\Delta_{p}$ and the magnon detuning $\Delta_{m}$. In this two-dimensional representation, the Fano resonances discussed in Fig.~\ref{Fig5} appear as characteristic ridge--valley structures: narrow high-transmission channels (constructive interference) adjacent to sharp suppression regions (destructive interference). The position, curvature, and contrast of these features encode how the interference between the broad cavity background and the narrow magnon--mechanical resonance is modified by the non-Hermitian coupling $\Gamma$. Since the Fano asymmetry parameter and linewidths are controlled by the effective complex susceptibilities of the magnon and mechanical modes, changing $\Gamma$ deforms the Fano contours in the $(\Delta_{p},\Delta_{m})$ plane and allows us to visualize directly how Hermitian MMIT gradually evolves into strongly asymmetric, gain-assisted Fano resonances.

Figure~\ref{Fig6}(a) shows the density plot of the normalized cavity transmission $T(\Delta_{p},\Delta_{m})$ in the Hermitian limit $\Gamma=0$ for fixed couplings $G_{a}/\Delta_{a}=2$ and $G_{b}/\Delta_{m}=0.1$. In this case, the pattern is nearly symmetric around the resonant line where the effective cavity and magnon detunings are matched, reflecting a conventional MMIT/Fano response dominated by coherent but loss-balanced interference. The high-transmission regions form smooth, slightly curved ridges, while the neighboring dark regions indicate frequencies and magnon detunings where destructive interference suppresses the probe. The absence of gain or excess loss ensures that the contrast between bright and dark regions is moderate and that the Fano profiles remain close to their Hermitian counterparts.
\begin{figure}[tp]
	\includegraphics[width=8.5cm]{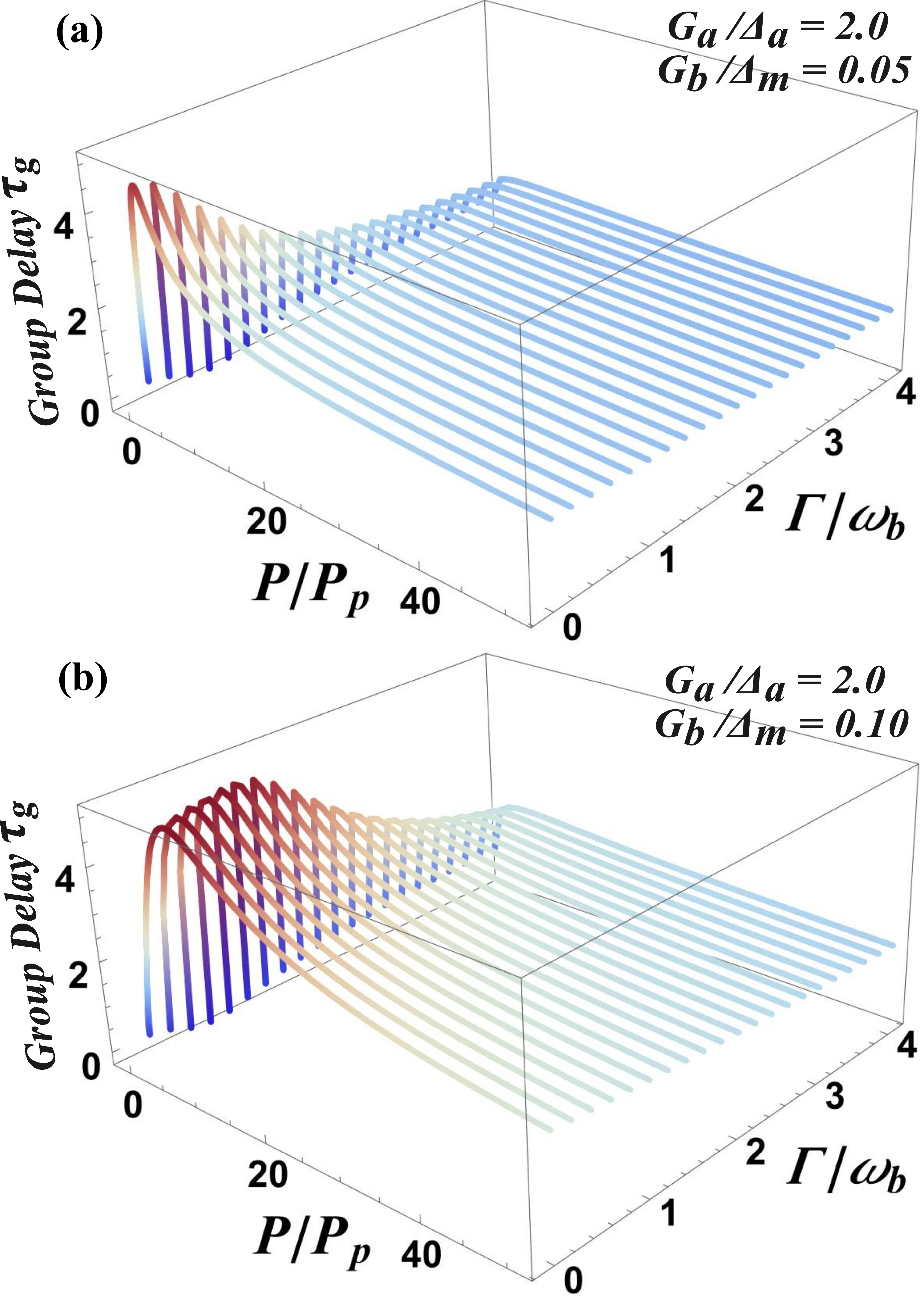}
	\caption{(Color online) Group delay $\tau_{g}$ of the probe field as a function of the normalized input power and non-Hermitian coupling $\Gamma/\kappa_{a}$ for (a) $G_{b}/\Delta_{m}=0.05$ and (b) $G_{b}/\Delta_{m}=0.1$, with fixed photon--magnon coupling $G_{a}/\Delta_{a}=2$. Regions with $\tau_{g}>0$ (slow light) and $\tau_{g}<0$ (fast light) demonstrate that the combined action of the non-Hermitian parameter and magnomechanical coupling provides wide tunability of the probe delay and advancement. The remaining parameters are the same as in Fig.~\ref{Fig2}.}
	\label{Fig7}
\end{figure}

In Figs.~\ref{Fig6}(b) and~\ref{Fig6}(c), we keep $G_{a}/\Delta_{a}=2$ and $G_{b}/\Delta_{m}=0.1$ fixed, but introduce and increase the non-Hermitian coupling to $\Gamma/\omega_{b}=2$ and $\Gamma/\omega_{b}=4$, respectively. For $\Gamma/\omega_{b}=2$ [Fig.~\ref{Fig6}(b)], the Fano structures become visibly distorted: the bright transmission ridges tilt and bend, and the contrast between the high- and low-transmission regions increases, particularly on the side where the gain-dominated hybrid mode contributes most strongly. This asymmetry in the density plot directly reflects the $\mathcal{PT}$-broken character of the hybrid eigenmodes in this regime, where one effective mode experiences net amplification while its partner is more strongly damped.

As $\Gamma/\omega_{b}$ is further increased to $4$ [Fig.~\ref{Fig6}(c)], the deformation becomes even more pronounced. The bright ridges of high transmission sharpen and can evolve into narrow, gain-assisted channels embedded in a more strongly absorbing background, while the dark regions become larger and deeper on the loss-dominated side of the parameter space. The resulting pattern exhibits strongly skewed, Fano-like contours, indicative of interference between an increasingly narrow, gain-enhanced resonance and the broader cavity continuum. Overall, Fig.~\ref{Fig6} shows that the non-Hermitian parameter $\Gamma$ provides a powerful means of engineering the topology and asymmetry of Fano resonances in the $(\Delta_{p},\Delta_{m})$ plane, enabling highly directional interference profiles, amplified transmission channels, and enhanced sensitivity to small changes in probe or magnon detuning.

\section{Fast and slow light across Exceptional Points}\label{sec4}
The control of fast and slow light plays a pivotal role in advancing quantum nonlinear optical processes toward practical implementations of quantum information processing and computation~\cite{Ref4,Ref5}. In the present system, the behavior of slow and fast light is governed by the phase $\Phi_{p}$ of the total probe transmission $\mathcal{E}_{p}$, which therefore serves as a key observable. In particular, the frequency dependence of $\Phi_{p}$ determines the probe group delay (or advancement), characterized by the group delay $\tau_{g}$, given by
\begin{eqnarray}
	\tau_g=&&\frac{\partial}{\partial\omega_p}\Phi_p(\omega_p)=\frac{\partial}{\partial\omega_p}\bigg(arg\big(\mathcal{T}_p(\omega_p)\big)\bigg)\nonumber\\
	=&&\frac{\partial}{\partial\omega_p}\bigg(arg\big(1-\sqrt{2\kappa}c_-(\Delta_p)/\eta_p\big)\bigg).\label{eq13}
\end{eqnarray}
\begin{figure*}[tp]
	\includegraphics[width=17cm]{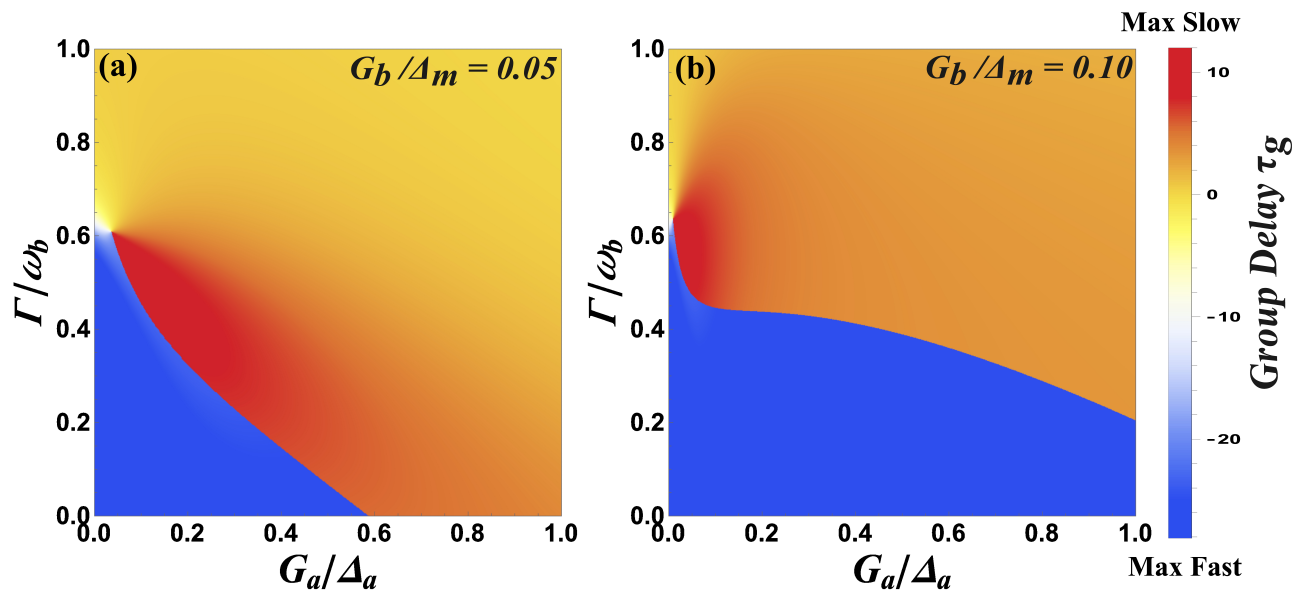}
	\caption{(Color online) Density plots of the probe group delay $\tau_{g}$ as functions of the photon--magnon coupling $G_{a}/\kappa_{a}$ and the non-Hermitian parameter $\Gamma/\kappa_{a}$ for (a) $G_{b}/\kappa_{a}=0.05$ and (b) $G_{b}/\kappa_{a}=0.1$. The color scale indicates the magnitude and sign of $\tau_{g}$, with positive (negative) values corresponding to slow-light (fast-light) propagation. Increasing $G_{b}/\kappa_{a}$ and tuning $(G_{a}/\kappa_{a},\Gamma/\kappa_{a})$ allows one to sculpt extended regions of large positive and negative group delay in parameter space. The remaining parameters are the same as in Fig.~\ref{Fig2}.}
	\label{Fig8}
\end{figure*}

Figure~\ref{Fig7} analyzes how the group delay of the probe field can be controlled by the input power and the non-Hermitian coupling $\Gamma$ in our cavity magnomechanical system. In each panel, the group delay $\tau_{g}$ is plotted as a function of the normalized input power for several values of $\Gamma/\kappa_{a}$, with the photon--magnon coupling fixed at $G_{a}/\kappa_{a}=2$. The two panels correspond to different magnomechanical couplings: $G_{b}/\kappa_{a}=0.05$ in Fig.~\ref{Fig7}(a) and $G_{b}/\kappa_{a}=0.1$ in Fig.~\ref{Fig7}(b). As the input power increases, the intracavity field grows and enhances the effective linearized couplings, thereby strengthening the interference responsible for MMIT and its non-Hermitian deformation. For small input power, the group delay remains modest because the transparency window and its associated phase slope are still weakly developed. As the power is raised, a pronounced region of positive group delay $\tau_{g}>0$ emerges, reflecting slow-light behavior associated with the steep normal dispersion in the (gain-assisted) MMIT window.

The presence of a finite non-Hermitian parameter $\Gamma$ qualitatively changes this picture. For moderate $\Gamma/\kappa_{a}$, the gain--loss imbalance in the hybrid magnon--photon subsystem increases the effective phase slope on one side of the resonance, leading to a substantial enhancement of the positive group delay over a finite range of input powers. As $\Gamma/\kappa_{a}$ is further increased, the system is driven deeper into the non-Hermitian (and $\mathcal{PT}$-broken) regime, and the phase dispersion can invert its curvature in the vicinity of the hybrid resonance. This inversion manifests as a crossover from positive to negative $\tau_{g}$, i.e., from slow to fast light, for appropriate combinations of input power and $\Gamma/\kappa_{a}$. In the fast-light regime, the transmitted probe pulse is advanced ($\tau_{g}<0$), a well-known consequence of anomalous dispersion in strongly interfering, gain-assisted media.

Comparing Figs.~\ref{Fig7}(a) and~\ref{Fig7}(b) shows the impact of the magnomechanical coupling on these slow-/fast-light effects. For $G_{b}/\kappa_{a}=0.05$, the mechanical contribution acts as a relatively weak perturbation: the regions of large positive or negative group delay are present but relatively narrow. When $G_{b}/\kappa_{a}$ is increased to $0.1$, the magnomechanical backaction becomes stronger and more strongly dresses the magnon susceptibility. As a result, the magnitude of both positive and negative group delays can be enhanced, and the parameter regions in which large $|\tau_{g}|$ is realized become more pronounced. Physically, this indicates that the mechanical mode, even when weakly coupled, provides an additional degree of freedom to tailor the dispersion of the hybrid system, thereby offering a route to engineer tunable slow- and fast-light behavior by jointly controlling the input power, the non-Hermitian parameter $\Gamma$, and the magnomechanical coupling $G_{b}$.

To obtain a more global view of how slow- and fast-light regimes are organized in parameter space, Fig.~\ref{Fig8} presents density plots of the probe group delay $\tau_{g}$ as functions of the photon--magnon coupling $G_{a}/\kappa_{a}$ and the non-Hermitian coupling $\Gamma/\kappa_{a}$. In both panels, other parameters are fixed as in Fig.~\ref{Fig2}, while the magnomechanical coupling is set to $G_{b}/\kappa_{a}=0.05$ in Fig.~\ref{Fig8}(a) and $G_{b}/\kappa_{a}=0.1$ in Fig.~\ref{Fig8}(b). The color scale encodes both the sign and magnitude of $\tau_{g}$, thereby distinguishing slow-light regions (positive group delay) from fast-light regions (negative group delay).

In Fig.~\ref{Fig8}(a), for weak magnomechanical coupling ($G_{b}/\kappa_{a}=0.05$), the parameter plane is divided into broad domains of slow and fast light, separated by contours where $\tau_{g}=0$. For small $\Gamma/\kappa_{a}$ and moderate $G_{a}/\kappa_{a}$, the system operates in a regime dominated by Hermitian-like MMIT, where the interference is mainly loss-balanced and the group delay is moderately positive. As $G_{a}/\kappa_{a}$ is increased, the photon--magnon hybridization strengthens, and the MMIT window becomes more pronounced, leading to an enhancement of the positive group delay over an extended region. By contrast, for larger $\Gamma/\kappa_{a}$, the non-Hermitian gain--loss imbalance reshapes the eigenmode spectrum. In this regime, portions of the parameter space exhibit negative $\tau_{g}$, corresponding to fast light, especially when $G_{a}/\kappa_{a}$ is large enough that the system resides near or beyond the $\mathcal{PT}$-symmetry-breaking threshold. The boundary between positive and negative group delay thus reflects the interplay between coherent coupling and non-Hermitian driving.

The effect of increasing the magnomechanical coupling is seen in Fig.~\ref{Fig8}(b), where $G_{b}/\kappa_{a}=0.1$. Here, the mechanical backaction more strongly modifies the effective magnon susceptibility, leading to a more structured distribution of $\tau_{g}$ in the $(G_{a}/\kappa_{a},\Gamma/\kappa_{a})$ plane. Regions of large positive group delay become more prominent for intermediate values of $G_{a}/\kappa_{a}$ and $\Gamma/\kappa_{a}$, corresponding to an optimal balance between strong MMIT-induced dispersion and non-Hermitian enhancement. At the same time, the fast-light regions with $\tau_{g}<0$ can extend over a wider range of parameters, especially in the high-$\Gamma$ sector where the gain-dominated hybrid mode exerts greater influence. The zero-delay contours separating slow and fast light acquire a more complex shape, reflecting the fact that the magnomechanical channel now significantly participates in the interference process.

Overall, Fig.~\ref{Fig8} demonstrates that the combined tuning of $G_{a}/\kappa_{a}$, $\Gamma/\kappa_{a}$, and $G_{b}/\kappa_{a}$ enables a high degree of control over the dispersion properties of the cavity magnomechanical system. By appropriately choosing these parameters, one can engineer extended regions of large positive group delay for slow-light applications, or large negative group delay for fast-light and advanced-pulse propagation, all within a single non-Hermitian MMIT platform.

\section{Conclusions}\label{sec5}
We have developed a comprehensive theory of magnomechanically induced transparency in a cavity magnomechanical platform with engineered non-Hermiticity and parity–time-symmetry–related amplification. Starting from the full linearized dynamics, we showed that, in the Hermitian limit, a strong photon–magnon coupling $G_{a}$ produces a single transparency window in the cavity response, which evolves into a double-window structure when the magnon is coherently hybridized with a mechanical mode via the magnomechanical coupling $G_{b}$. Although $G_{b}\ll G_{a}$ in typical experiments, we demonstrated that scaled values of $G_{b}/\Delta_{m}$ can drive a clear splitting of the transparency window, and we discussed how such regimes can be approached experimentally by optimizing the YIG size, mechanical quality factor, and magnon–phonon mode overlap.

Introducing a traveling-field–induced non-Hermitian coupling $\Gamma$ transforms conventional MMIT into a gain-assisted, $\mathcal{PT}$-symmetry–related phenomenon. In this regime, the hybrid magnon–photon eigenmodes acquire imbalanced effective linewidths, and the transparency feature becomes asymmetric, with suppressed transmission on the loss-dominated side and $T(\Delta_{p})>1$ amplification on the gain-dominated side. We linked this behavior to the proximity of exceptional-point physics identified in the effective eigenvalue spectrum. We further showed that, in the presence of finite $\Gamma$ and weak $G_{b}$, tuning the cavity detuning $\Delta_{a}$ sculpts the MMIT feature into Fano-like profiles with pronounced peak–dip asymmetry and strongly non-Lorentzian phase dispersion. Two-dimensional density plots in the $(\Delta_{p},\Delta_{m})$ plane revealed how increasing $\Gamma$ continuously deforms nearly symmetric Hermitian interference patterns into highly directional, gain-assisted Fano ridges.

Finally, we analyzed the group delay associated with the transmitted probe and demonstrated that both slow and fast light can be realized and widely tuned by adjusting $G_{a}$, $G_{b}$, and $\Gamma$. The non-Hermitian control of MMIT, Fano resonances, and group delay established here shows that cavity magnomechanical systems provide a powerful and experimentally feasible platform for implementing reconfigurable, $\mathcal{PT}$-symmetry–based quantum transducers, amplifiers, and precision interferometric sensors using engineered gain–loss and magnomechanical interference.

\begin{acknowledgments}
W.M.L. acknowledges the support from National Key R\&D Program of China under grants No. 2021YFA1400900, 2021YFA0718300, 2021YFA1402100, NSFC under Grants Nos. 12174461, 12234012, 12334012, 52327808, Space Application System of China Manned Space Program. K.A.Y. acknowledges the support of Research Fund for International Young Scientists by NSFC under grant No. KYZ04Y22050, Zhejiang Normal University research funding under grant No. ZC304021914 and Zhejiang province postdoctoral research project under grant number ZC304021952. 
\end{acknowledgments}

\section*{Data Availability}

The data that support the findings of this study are available within the article.

\appendix
\section{Linear stability analysis}\label{App:Stability}

In this Appendix we derive the linear stability conditions of the non-Hermitian cavity magnomechanical system using the Routh–Hurwitz criterion. We start from the linearized quantum Langevin equations for the cavity, magnon, and mechanical fluctuations (noise terms omitted for clarity),
\begin{eqnarray}
	\delta \dot{\hat{a}} & = & -(i\Delta_a+\kappa_a)\,\delta \hat{a}
	-(iG_a+\Gamma e^{i\theta})\,\delta \hat{m}, \\
	\delta \dot{\hat{m}} & = & -(i\Delta_m+\kappa_m)\,\delta \hat{m}
	- iG_{mb}\,\delta \hat{q}
	-(iG_a+\Gamma e^{i\theta})\,\delta \hat{a}, \\
	\delta \dot{\hat{p}} & = & -\omega_b\,\delta \hat{q}
	+G_b\left(\delta \hat{m} + \delta \hat{m}^\dagger\right)
	-\gamma_b\,\delta \hat{p}, \\
	\delta \dot{\hat{q}} & = & \omega_b\,\delta \hat{p}.
\end{eqnarray}
To obtain a real drift matrix, we introduce the quadrature operators
\begin{eqnarray}
	\delta \hat{X}_a &=& \frac{\delta \hat{a} + \delta \hat{a}^\dagger}{\sqrt{2}}, \qquad
	\delta \hat{Y}_a = \frac{\delta \hat{a} - \delta \hat{a}^\dagger}{i\sqrt{2}}, \\
	\delta \hat{X}_m &=& \frac{\delta \hat{m} + \delta \hat{m}^\dagger}{\sqrt{2}}, \qquad
	\delta \hat{Y}_m = \frac{\delta \hat{m} - \delta \hat{m}^\dagger}{i\sqrt{2}}.
\end{eqnarray}
In terms of these quadratures, the fluctuation vector
\begin{equation}
	\mathbf{u}(t) =
	\bigl(
	\delta X_a, \,
	\delta Y_a, \,
	\delta X_m, \,
	\delta Y_m, \,
	\delta q, \,
	\delta p
	\bigr)^{\mathrm{T}}
\end{equation}
obeys a linear differential equation of the form
\begin{equation}
	\dot{\mathbf{u}}(t) = A\,\mathbf{u}(t) + \mathbf{n}(t),
\end{equation}
where $\mathbf{n}(t)$ collects the corresponding noise operators and
$A$ is the $6\times 6$ drift matrix. Writing the complex cavity–magnon coupling as
\begin{equation}
	iG_a + \Gamma e^{i\theta}
	= K_{\mathrm{r}} + iK_{\mathrm{i}},
	\qquad
	K_{\mathrm{r}} = \Gamma\cos\theta,
	\qquad
	K_{\mathrm{i}} = G_a + \Gamma\sin\theta,
\end{equation}
the explicit form of $A$ reads
\begin{widetext}
	\begin{equation}
		A =
		\begin{pmatrix}
			-\kappa_a & \Delta_a   & -K_{\mathrm{r}} &  K_{\mathrm{i}} &     0      &     0      \\
			-\Delta_a & -\kappa_a  & -K_{\mathrm{i}} & -K_{\mathrm{r}} &     0      &     0      \\
			-K_{\mathrm{r}} &  K_{\mathrm{i}} & -\kappa_m &  \Delta_m   &     0      &     0      \\
			-K_{\mathrm{i}} & -K_{\mathrm{r}} & -\Delta_m & -\kappa_m   & -G_{mb}    &     0      \\
			0          &     0      &     0      &     0      &     0      &  \omega_b \\
			0          &     0      & \sqrt{2}\,G_b &  0      & -\omega_b  & -\gamma_b
		\end{pmatrix}.
		\label{A-matrix}
	\end{equation}
\end{widetext}
The linearized dynamics is stable if and only if all eigenvalues of the drift matrix $A$ have strictly negative real parts. Equivalently, stability requires that the characteristic polynomial of $A$,
\begin{equation}
	\det(\lambda I - A)
	=
	\lambda^{6} + s_{1}\lambda^{5} + s_{2}\lambda^{4}
	+ s_{3}\lambda^{3} + s_{4}\lambda^{2} + s_{5}\lambda + s_{6},
	\label{char-poly}
\end{equation}
has all roots with $\mathrm{Re}\,\lambda < 0$. Here the real coefficients
$s_{j} = s_{j}(\kappa_{a},\kappa_{m},\gamma_{b},\Delta_{a},\Delta_{m},
\omega_{b},G_{a},G_{b},G_{mb},\Gamma,\theta)$ are determined by the entries
of the drift matrix $A$ in Eq.~(\ref{A-matrix}). In particular,
$s_{1}=-\mathrm{Tr}\,A>0$, $s_{6}=\det(-A)>0$, and the remaining
$s_{2},\dots,s_{5}$ can be expressed in terms of traces of powers of $A$
and principal minors, although their explicit analytic forms are lengthy
and not reproduced here.

According to the Routh--Hurwitz criterion for a sixth-order polynomial
of the form~(\ref{char-poly}), a necessary and sufficient condition for
all roots to have negative real parts is that
\begin{equation}
	s_{j} > 0 \quad (j=1,\dots,6),
\end{equation}
and that all leading principal Hurwitz determinants $\Delta_{k}$ $(k=1,\dots,6)$
built from the coefficients $\{s_{j}\}$ are positive. The first few
determinants can be written explicitly as
\begin{eqnarray}
	\Delta_{1} &=& s_{1}, \\
	\Delta_{2} &=& 
	\begin{vmatrix}
		s_{1} & 1 \\
		s_{3} & s_{2}
	\end{vmatrix}
	= s_{1}s_{2} - s_{3}, \\
	\Delta_{3} &=&
	\begin{vmatrix}
		s_{1} & 1   & 0 \\
		s_{3} & s_{2} & s_{1} \\
		s_{5} & s_{4} & s_{3}
	\end{vmatrix}
	= s_{1}s_{2}s_{3} - s_{1}^{2}s_{4} - s_{3}^{2} + s_{1}s_{5},
\end{eqnarray}
and the higher-order determinants $\Delta_{4}$, $\Delta_{5}$, and
$\Delta_{6}$ follow from the standard Hurwitz construction for degree-six polynomials.
The full set of Routh--Hurwitz stability conditions therefore reads
\begin{equation}
	s_{j} > 0 \quad (j=1,\dots,6),
	\qquad
	\Delta_{k} > 0 \quad (k=1,\dots,6).
	\label{RH-conditions}
\end{equation}

Equations~(\ref{A-matrix})--(\ref{RH-conditions}) provide the stability
criteria for the non-Hermitian cavity magnomechanical system in terms of
its microscopic parameters. In practice, one computes the coefficients
$s_{j}$ from the drift matrix $A$ for a given set of parameters and
then evaluates the Hurwitz determinants $\Delta_{k}$. Only parameter
regimes satisfying all inequalities in Eq.~(\ref{RH-conditions}) are
used in the main text, ensuring that the steady state about which the
system is linearized is dynamically stable.


\begin{thebibliography}{99}
	
	\bibitem{Aspelmeyer2014}
	M. Aspelmeyer, T. J. Kippenberg, and F. Marquardt,
	Rev. Mod. Phys. \textbf{86}, 1391 (2014).
	
	\bibitem{Zhang2016_SciAdv}
	X. Zhang, C.-L. Zou, L. Jiang, and H. X. Tang,
	Sci. Adv. \textbf{2}, e1501286 (2016).
	
	\bibitem{Zhang2016}
	D. Zhang, X.-Q. Luo, Y.-P. Wang, T.-F. Li, and J. Q. You,
	npj Quantum Inf. \textbf{1}, 15014 (2016).
	
	\bibitem{Fan2022}
	Z.-Y. Fan, H. Qian, and J. Li,
	Quantum Sci. Technol. \textbf{8}, 015014 (2022).
	
	\bibitem{Hu2013}
	H. Huebl, C. W. Zollitsch, J. Lotze, \textit{et al.},
	Phys. Rev. Lett. \textbf{111}, 127003 (2013).
	
	\bibitem{Tabuchi2015}
	Y. Tabuchi, S. Ishino, A. Noguchi, \textit{et al.},
	Science \textbf{349}, 405 (2015).
	
	\bibitem{Harder2021}
	M. Harder and C.-M. Hu,
	Phys. Rep. \textbf{905}, 1 (2021).
	
	\bibitem{Wang2018}
	Y.-P. Wang, G.-Q. Zhang, D. Zhang, T.-F. Li, C.-M. Hu, and J. Q. You,
	Phys. Rev. A \textbf{97}, 013821 (2018).
	
	\bibitem{Rao2020}
	J. W. Rao, C. Yu, and G. Li,
	Phys. Rev. B \textbf{101}, 064423 (2020).
	
	\bibitem{Luo2023}
	Y.-X. Luo, L.-J. Cong, Z.-G. Zheng, \textit{et al.},
	Opt. Express \textbf{31}, 34764 (2023).
	
	\bibitem{ElGanainy2018}
	R. El-Ganainy, K. G. Makris, M. Khajavikhan, \textit{et al.},
	Nat. Phys. \textbf{14}, 11 (2018).
	
	\bibitem{Feng2017}
	L. Feng, R. El-Ganainy, and L. Ge,
	Nat. Photonics \textbf{11}, 752 (2017).
	
	\bibitem{Ozdemir2019}
	\c{S}. K. \"{O}zdemir, S. Rotter, F. Nori, and L. Yang,
	Nat. Mater. \textbf{18}, 783 (2019).
	
	\bibitem{Bender1998}
	C. M. Bender and S. Boettcher,
	Phys. Rev. Lett. \textbf{80}, 5243 (1998).
	
	\bibitem{Miri2019}
	M.-A. Miri and A. Al\`u,
	Science \textbf{363}, eaar7709 (2019).
	
	\bibitem{Peng2014}
	B. Peng, \c{S}. K. \"{O}zdemir, F. Lei, \textit{et al.},
	Nat. Phys. \textbf{10}, 394 (2014).
	
	\bibitem{Zhang2019}
	D. Zhang, X.-Q. Luo, Y.-P. Wang, T.-F. Li, and J. Q. You,
	Nat. Commun. \textbf{10}, 1368 (2019).
	
	\bibitem{Qian2024}
	J. Qian, J. Li, S.-Y. Zhu, J. Q. You, and Y.-P. Wang,
	Phys. Rev. Lett. \textbf{132}, 156901 (2024).
	
	\bibitem{Jing2014}
	H. Jing, \c{S}. K. \"{O}zdemir, X.-Y. L\"u, J. Zhang, L. Yang, and F. Nori,
	Phys. Rev. Lett. \textbf{113}, 053604 (2014).
	
	\bibitem{Zhang2020}
	Z. Zhang, Y.-P. Wang, and X. Wang,
	Phys. Rev. A \textbf{102}, 023512 (2020).
	
	\bibitem{Dai2024}
	T. Dai, Y. Ao, J. Mao, \textit{et al.},
	Nat. Phys. \textbf{20}, 101 (2024).
	
	\bibitem{Chengyong2025}
	Y. Chengyong, W.-M. Liu, and K. A. Yasir, npj Quantum Mater. \textbf{10}, 108 (2025).
	
	\bibitem{Lai2024}
	C. Lai, S. Fahad, and K. A. Yasir,
	Results Phys. \textbf{64}, 107917 (2024).
	
	\bibitem{Fleischhauer2005}
	M. Fleischhauer, A. Imamoglu, and J. P. Marangos,
	Rev. Mod. Phys. \textbf{77}, 633 (2005).
	
	\bibitem{Weis2010}
	S. Weis, R. Rivi\`ere, S. Del\'eglise, \textit{et al.},
	Science \textbf{330}, 1520 (2010).
	
	\bibitem{Safavi2011}
	A. H. Safavi-Naeini, T. P. Mayer Alegre, J. Chan, \textit{et al.},
	Nature \textbf{472}, 69 (2011).
	
	\bibitem{Agarwal2010}
	G. S. Agarwal and S. Huang,
	Phys. Rev. A \textbf{81}, 041803(R) (2010).
	
	\bibitem{Xu2022}
	J. Xu, R. Liu, and L. Ge,
	Phys. Rev. A \textbf{106}, 033502 (2022).
	
	\bibitem{He2019}
	H. He, M. C. Kuzyk, J. Ren, \textit{et al.},
	Phys. Rev. A \textbf{100}, 023820 (2019).
	
	\bibitem{Zhang2021}
	Y. Zhang, X. Shen, and J. Li,
	Phys. Rev. A \textbf{104}, 033501 (2021).
	
	\bibitem{Yu2023}
	H. Yu, Z. Qiu, and Y. Liu,
	Optica \textbf{10}, 193 (2023).
	
	\bibitem{Yang2023}
	F. Yang, Z. Zhou, and Y. Li,
	Phys. Rev. Lett. \textbf{131}, 243602 (2023).
	
	\bibitem{Ren2024}
	Q. Ren, C. Hu, and X. Li,
	Phys. Rev. Appl. \textbf{21}, 054009 (2024).
	
	\bibitem{Qiu2023}
	L. Qiu, Y. Liu, and S. Gong,
	Nat. Commun. \textbf{14}, 4253 (2023).
	
	\bibitem{Yasir2022}
	K. A. Yasir, L. Zhuang, and W.-M. Liu,
	npj Quantum Inf. \textbf{8}, 109 (2022).
	
	\bibitem{Yasir2017}
	K. A. Yasir, L. Zhuang, and W.-M. Liu,
	Phys. Rev. A \textbf{95}, 013810 (2017).
	
	\bibitem{Yasir2016}
	K. A. Yasir and W.-M. Liu,
	Sci. Rep. \textbf{6}, 22651 (2016).
	
	\bibitem{Yasir2023}
	K. A. Yasir, Z. Liang, and W.-M. Liu,
	Eur. Phys. J. Plus \textbf{138}, 29 (2023).
	
	\bibitem{Noori2014}
	B. Peng, Ş. K. Özdemir, F. Lei, F. Monifi, M. Gianfreda, G. L. Long, S. Fan, F. Nori, C. M. Bender, and L. Yang,
	Nat. Phys. \textbf{10}, 394 (2014).
	
	\bibitem{Noori2019}
	Ş. K. Özdemir, S. Rotter, F. Nori, and L. Yang,
	Nat. Mater. \textbf{18}, 783 (2019).
	
	\bibitem{Noori2021}
	H. Xu, D.-G. Lai, Y.-B. Qian, B.-P. Hou, A. Miranowicz, and F. Nori,
	Phys. Rev. A \textbf{104}, 053518 (2021).
	
	\bibitem{Ref280}J. F. Scott, \textit{Rev. Mod. Phys.} \textbf{46}, 83 (1974).
	\bibitem{Ref2800}A. E. Miroshnichenko, S. Flach, and Y. S. Kivshar \textit{Rev. Mod. Phys.} \textbf{82}, 2257 (2010).
	
	\bibitem{Ref4}H. J. Kimble, \textit{Nature} \textbf{453}, 1023-1030 (2008).
	\bibitem{Ref5}A. I. Lvovksy, B. C. Sanders, and W. Tittel, \textit{Nat. Photon.} \textbf{3}, 706-714 (2009).
	
\end{thebibliography}
\end{document}